\documentclass{proceedingsM}
\usepackage{times,amsmath,amssymb}
\newcommand{\pp}{\partial}

\title{On modelling of physical effects accompanying 
the propagation of action potentials in nerve fibres}
\author{J\"uri Engelbrecht, Tanel Peets, Kert Tamm, Martin Laasmaa, Marko Vendelin}
\address{Centre for Nonlinear Studies, Institute of Cybernetics at TUT, Akadeemia tee 21, Tallinn 12618, Estonia}
\email{je@ioc.ee, tanelp@ioc.ee, kert@ioc.ee, martin@sysbio.ioc.ee, markov@sysbio.ioc.ee}

\abstract{The recent theoretical and experimental studies have revealed many details of signal propagation in nervous systems. In this paper an attempt is made to unify various mathematical models which describe the signal propagation in nerve fibres. The analysis of existing single models permits to select the leading physiological effects. As a result, a more general mathematical model is described based on the coupling of action potentials with mechanical waves in a nerve fibre. The crucial issue is how to model coupling effects which are strongly linked to the ion currents through biomembranes.
}
\keywords{nerve fibres, action potentials, mechanical waves, biomembranes}

\begin{document}

\maketitle
\section{Introduction}

The modelling of nerve pulse propagation is historically related to the analysis of action potentials which have been measured using electrophysical techniques. The celebrated Hodgkin-Huxley (HH) model \cite{Hodgkin1952} describes the propagation of asymmetric pulse in a nerve fibre which can be considered as a cylindrical biomembrane filled with axoplasm. Due to ion currents (mostly Na and K) through the membrane this pulse is able to propagate with a constant shape with an overshoot responsible for the refraction time. Hodgkin and Huxley \cite{Hodgkin1952} have proposed kinetic equations for governing ion currents and have established experimentally the needed parameters. In order to grasp mathematically the essence of the nerve pulse propagation, several simpler models have been proposed (FitzHugh-Nagumo \cite{FitzHugh1961}; Engelbrecht \cite{Engelbrecht1981}, etc). A recent study \cite{Bresslof2014} has summarised the present knowledge on mathematical modelling of action potentials including neural networks. In parallel to studies of action potentials, it has been shown experimentally that mechanical displacements and heat transfer in axonal membrane accompany the electrical signal \cite{Tasaki1988,Iwasa1980}. In recent years these studies have been intensified thanks to the growing interest to the behaviour of biomembranes which are the main building blocks of biology \cite{Mueller2014}.

The starting point in modelling is the description of an electrical signal (action potential) which propagates in axoplasm surrounded by a cylindrical biomembrane and is understood as a change in the potential within the axoplasm. This cylindrical biomembrane is surrounded from outside by the intersticial fluid. The action potential has a constant shape which is supported by ion currents through the biomembrane. The well-known models for action potentials like HH, FitzHugh-Nagumo (FHN) or evolution equations \cite{Engelbrecht1981} include mechanisms of ion currents which according to the terminology of HH, are phenomenological variables. From the viewpoint of thermodynamics, however, these can be considered as internal variables \cite{Maugin1994}.

Next, the processes in biomembranes must be understood. Much attention is paid to the fundamentals of biomembranes which play an important role in many cellular processes including propagation of nerve impulses (overview by Mueller and Tyler \cite{Mueller2014}). Two main problems can be distinguished: mechanisms of opening ion channels and mechanisms of mechanical behaviour. The structure of biomembranes is nowadays well understood: they are made of phospholipids \cite{Mueller2014,Tieleman2003} in the form of bilayers which can exist either in a liquid phase or in a gel phase. Such a membrane is able to deform because of its elastic properties \cite{Barz2013} in the longitudinal and transverse direction as well as to undergo bending \cite{Mueller2014}. The studies involve explanation of dynamical deformation of membranes \cite{Heimburg1998}.

The next important feature related to the action potentials is that it generates an axoplasmic pressure pulse propagating synchronously with the action potential \cite{Rvachev2010}. Such a pressure pulse will die out without the constant support from an action potential as shown by Rvachev \cite{Rvachev2010}.

Finally, the closing link to the full description of the physiological phenomena in nerve fibres is the impact of both the action potential and the pressure pulse on the membrane which should explain opening the ion channels, the deformation of membranes, the generation of heat and accompanying optical effects \cite{Heimburg1998,Mueller2014,Andersen2009}.

An important element of signal propagation in neural systems beside the propagation in fibres is the transmittance of the signals across synaptic contacts to receptors (terminals). This transmittance is realised by chemical and/or electrical signals \cite{Debanne2011,Hormuzdi2004} and the electrical signalling may be interrelated with chemical transmission \cite{Hormuzdi2004}. Recent experiments have shown that a mechanical spike may also accompany the arrival of the action potential to the terminal volume \cite{Kim2007}. The role of a mechanical spike beside electrical and chemical transmittance is not clear yet. However, as said by Bennet \cite{Bennett2000}: `the nervous system does its operations in a number of different ways'. 

It is a challenge to understand coupling between action potential and the accompanying effects together with possible structural changes in biomembranes. Up to now, there is no theoretical consensus about the general models which could describe all the measured phenomena within one framework. Clearly the structural properties of biomembranes \cite{Mueller2014} must be taken into account. Beside models based on electrophysics, the models are proposed which prescribe the signal transfer to mechanics only \cite{Heimburg1998,Andersen2009} or to protein thermodynamic structures \cite{Zhao2013}. Here in this paper we sum up the previous results (Section 2) and propose to unite the various models into a more plausible description based on coupling of various effects (Section 3). The brief discussion follows in Section 4. 

\section{Existing physical models and their parameters}

\subsection{Action potential}
Both the axoplasm and the intersticial fluid contain ions of sodium (Na+) and potassium (K+) as well as other ions \cite{Hodgkin1952,Scott1999}. The relative concentration of ions create the transmembrane potential which at the equilibrium value is in the range of -50 to -100 millivolts. If a stimulus applied to a nerve is below a certain threshold value then the equilibrium value will be restored fast. If a stimulus is large enough (above the threshold), then a stable action potential will be formed and it starts to propagate along the nerve fibre. For a standard HH axon \cite{Scott1999} with a diameter about 0.5 mm, the amplitude is about +50 millivolts and it has a typical overshoot of about 20 millivolts. The whole duration of a pulse is about 4 -6 msec and the propagation velocity is about 20 m/sec or more. The Hodgkin-Huxley (HH) model which describes the propagation of such a pulse is the following \cite{Hodgkin1952}:
(i) the basic telegraph equation where the inductance is neglected (a diffusion-type equation):
\begin{equation}
\label{eq1}
\frac{\pp^2 v}{\pp x^2}=cr_s\frac{\pp v}{\pp t} + r_s I_i
\end{equation}
where $v$ is the voltage, $c$ - the capacitance, $r_s$ the resistance determined  as $1/(\pi a^2 \sigma_1)$, $a$ is the radius of an axon and $\sigma_1$ - the conductivity and $I_i$ is the ion current. 

(ii) a phenomenological expression for the ion current:
\begin{equation}
\label{eq2}
I_i=g_Kn^4(v-v_R-v_K)+g_{Na}m^3h(v-v_R-v_{Na})+g_L(v-v_R-v_L)+C_m\frac{\pp v}{\pp t},
\end{equation}
\begin{equation}
\label{eq3}
g_K=2\pi a G_K, \quad g_{Na}=2\pi a G_{Na}, \quad g_L=2\pi a G_L.
\end{equation}
Here $G_K$ and $G_{Na}$ are the maximum potassium and sodium conductances, respectively; $G_L$ is a constant leakage conductance, $C_m$ is the membrane capacitance per unit area, $v_R$ is the resting potential and $v_K$, $v_{Na}$, $v_L$ are the corresponding equilibrium potentials. The phenomenological (hidden) variables $n$, $m$, and $h$ govern the "turning on" and "turning off" the membrane conductances. 

(iii) kinetic equations for three phenomenological variables:
\begin{subequations}
\begin{align}
&\frac{dn}{dt}=\alpha_n(1-n)-\beta_n n,  \\
&\frac{dm}{dt}=\alpha_m(1-m)-\beta_m m, \\
&\frac{dh}{dt}=\alpha_h(1-h)-\beta_h h,
\end{align}
\end{subequations}
where $\alpha_i$, $\beta_i$ are related to corresponding equilibrium values and relaxation times.

(iv) expressions for the parameters of the kinetic equations:
\begin{subequations}
\begin{align}
&\alpha=0.01(10-v)[\exp (10-v)/10-1]^{-1}, \\
&\beta_n=0.125\exp (-v/80),\\
&\alpha_m=0.01(25-v)[\exp (25-v)/10-1]^{-1},\\
&\beta_m=4\exp (-v/18),\\
&\alpha_h=0.07\exp (-v/20),\\
&\beta_h=[\exp (30-v)/10+1]^{-1}.
\end{align}
\end{subequations}
determined experimentally \cite{Hodgkin1952}.

The simplified FitzHugh-Nagumo (FHG) model takes into account only one phenomenological variable and reads \cite{Nagumo1962}
\begin{equation}
\label{eq6}
\frac{\pp^3 z}{\pp t \pp x^2}=\frac{\pp^2 z}{\pp t^2}+\mu (1-z+\epsilon z^2)\frac{\pp z}{\pp t}+z,
\end{equation}
where $\mu$ and $\epsilon$ are parameters and $z$ is the scaled voltage. 

One can add to these traditional widely used models an evolution equation derived in \cite{Engelbrecht1981}
\begin{equation}
\label{eq7}
\frac{\pp^2 z_1}{\pp x \pp \xi}+f(z_1)\frac{\pp z_1}{\pp \xi}+g(z_1)=0,
\end{equation}
where $z_1$ is the scaled voltage and $\xi$ is a moving frame $\xi=c_0t-x$. Following the FHN model with one phenomenological variable, $f(z)$ is a quadratic function and $g(z)$ - a linear function. Note that the moving frame involves velocity $c_0$ determined from the full telegraph equation. The final velocity $c$ of a nerve pulse is different from $c_0$. 

It must, however, be stressed that the ionic mechanisms are in fact more complicated than assumed even in the HH model with three variables and include several pump, exchange and background currents \cite{Courtemanche1998}. The diversity of channels is reflected also in a wide range of shapes and patterns for nerve pulses \cite{Bean2007}.

The models described above are able to reflect the main features of measured action potentials: (i) the existence of a threshold for an input; (ii) the all-or-non phenomenon for a pulse; (iii) the existence of an asymmetric localised pulse with an overshoot; (iv) the existence of a refraction length; (v) the possible annihilation of counter-propagating pulses \cite{Hodgkin1945,Nagumo1962}. The propagation of an action potential is triggered by an electrical excitation. A recent overview on processes in axons from the physiological viewpoint is given in \cite{Debanne2011}. Despite of such clear and measurable characteristics of pulses, the governing equations of action potentials can not describe other accompanying effects like mechanical displacements in the fibre walls or heat transfers -- see the recent criticism about the HH model by \cite{Appali2012a}. 

\subsection{Biomembranes}
Lipid membranes are not only a part of nerve fibres but they are important constituents of cells in much wider context \cite{Mueller2014}. The behaviour of membranes under mechanical forces is experimentally observed. In the equilibrium condition the membrane is in a liquid phase but under mechanical excitation the membrane undergoes a structural change from the liquid phase to the gel phase \cite{Heimburg1998,Andersen2009}. Consequently, its compressibility will be changed, too. The measurements of compressibility are carried out by calorimetry \cite{Heimburg1998,Schrader2002} and by ultrasonic velocity measurements \cite{Schrader2002}. This is needed for building up the dynamical models for waves in membranes like described in \cite{Heimburg2005} where the energy density is given by a nonlinear expression:
\begin{equation}
\label{eq8}
e=\frac{c_0^2}{\rho_0}u^2+\frac{p}{3\rho_0}u^3+\frac{q}{6\rho_0}u^4,
\end{equation}
where $c_0$ is the velocity of the longitudinal wave in a membrane, $\rho_0$ is the density and $u=\Delta\rho$ is the amplitude of the change in the membrane area while $p$ and $q$ are constants. 

The structural changes in membranes under forcing are not only related to phase transformation, these are also related to opening the channels (pores) for ions \cite{Mueller2014,Tieleman2003} and for water molecules \cite{Barz2013}. Such mechanosensitive changes play an important role for ion currents and can explain the role of anesthetics which may depend on the expansion modulus of membranes \cite{Rvachev2010}. The opening of channels under mechanical stress and electric field is demonstrated also by molecular dynamics simulations \cite{Tieleman2003,Shillcock2002,Bockmann2008}.
\subsection{Dynamics of membranes}
As explained by \cite{Mueller2014}, biomembranes are able to resist pressure, tension, stretch and bending. The deformation can be induced by mechanical or electrical impact \cite{Griesbauer2012,Shrivastava2014,Petrov2001}. The pressure pulse have been studied theoretically \cite{Griesbauer2012} and experimentally \cite{Shrivastava2014}. The theoretical model of Griesbauer et al \cite{Griesbauer2012} for pulses in lipid monolayers at the air-water interface is a second-order wave equation which is viscously coupled (the first derivative) to the liquid (water) underneath. The existence of 2D sound waves along a lipid monolayer is demonstrated by Shrivastava and Schneider \cite{Shrivastava2014}. The localized self-supporting pulses have been found having a threshold amplitude and all-or-none nature. The need to take nonlinearity in compressibility together with dispersive effects into account is stressed. 

Such a nonlinear model with dispersion is proposed by Heimburg and Jackson \cite{Heimburg2005}, Andersen et al \cite{Andersen2009}, etc. The governing equation for a 1D pulse in a cylindrical biomembrane is proposed by using the energy density \eqref{eq8} resulting in 
\begin{equation}
\label{eq9}
\frac{\pp^2 u}{\pp t^2}=\frac{\pp }{\pp x}\left[\left(c_0^2+pu+qu^2\right)\frac{\pp u}{\pp x}\right]-h\frac{\pp^4u}{\pp x^4},
\end{equation}
where $u$ is the density change (cf. \eqref{eq8}) and $h$ is an \textit{ad hoc} constant. This is a Boussinesq-type equation where nonlinearity is caused by the compressibility which has an impact on the velocity:  
\begin{equation}
\label{eq10}
c^2=c_0^2+pu+qu^2.
\end{equation}
Due to the existence of nonlinearities and dispersion in Eq. \eqref{eq8}, the solution of Eq. \eqref{eq8} may have a solitary character \cite{Heimburg2005}. In order to improve dispersive properties of such a model, it has been proposed to use more realistic dispersion terms like it is done for modelling waves in rods \cite{Engelbrecht2015}:
\begin{equation}
\label{eq11}
\frac{\pp^2 u}{\pp t^2}=\frac{\pp }{\pp x}\left[\left(c_0^2+pu+qu^2\right)\frac{\pp u}{\pp x}\right]-h_1\frac{\pp^4u}{\pp x^4}+h_2\frac{\pp^4u}{\pp x^2\pp t^2}
\end{equation}
where $h_1$, $h_2$ are dispersive constants. 

In order to compare these theoretical results with experiments \cite{Tasaki1988} one has to understand that the Eqs. \eqref{eq9}, \eqref{eq10} describe longitudinal waves and the transverse displacements $w$ measured by \cite{Tasaki1988} can be found from the derivative of the longitudinal profile \cite{Engelbrecht2015,Hady2015}:
\begin{equation}
\label{eq12}
w=-kr\frac{\pp u}{\pp x},
\end{equation}
where $r$ is the radius of the axon and $k$ is a constant. In the theory of rods $k$ is the Poisson ratio. However, it is unclear how fast the longitudinal waves in biomembranes decay without amplification.

The wave equations \eqref{eq9} and \eqref{eq11} density waves in biomembranes and are able to model localised pulses for given nonlinearities and dispersive terms. As far as these localised pulses are of a solitary character, it is proposed to call such models ` a soliton model' for signals in nerve fibres \cite{Heimburg1998,Andersen2009,Heimburg2005,Appali2012}. It should be stressed that although such a signal corresponds to changes in a biomembrane from a liquid to a gel state \cite{Mueller2014, Heimburg2005}, in its essence it is a longitudinal wave and the transverse displacement (swelling of a fibre) must be calculated according to expression \eqref{eq12}. It is shown that nonlinear equations \eqref{eq9} and \eqref{eq11} possess soliton-type solution \cite{Heimburg2005,Lautrup2011,Engelbrecht2015,Peets2015a}. Compared with the action potential, a soliton has not got an overshoot. It is argued that higher density in a soliton should be accompanied by a dilated region due to the mass conservation and this should prevent close sequence of pulses \cite{Gonzalez-Perez2014}. Intuitively it is understood but there is no evidence about such a phenomenon in modelling. According to \cite{Heimburg2005,Lautrup2011}, a soliton is formed by an initial input \cite{Lautrup2011} that is a pressure change. It is stressed that the  mechanical wave (soliton) is associated with a transient voltage change which affects the membrane potential difference \cite{Appali2012a} or, in other words `the appearance of a voltage pulse is merely a consequence of the piezo-electric nature of the nerve membrane which is partially charged and asymmetric' \cite{Lautrup2011}. However, this mechanism seemingly needs more detailed inspection. A good side of model \eqref{eq9} is that it allows the collision of pulses \cite{Lautrup2011} which is experimentally observed in giant axons of the earthworm \cite{Gonzalez-Perez2014}.

From a viewpoint of mathematical physics, eqs \eqref{eq9} and \eqref{eq11} are of the Boussinesq type \cite{Christov2007} reflecting nonlinear and dispersive effects. However, the nonlinearity in terms of $u$ in wave equations is of a different character from the usual $u_x$  type. The soliton solutions to such equations are found \cite{Heimburg2005,Lautrup2011,Peets2015a} and the emergence of soliton trains described \cite{Engelbrecht2015,Peets2015a}. In the emergence process the influence of a specific nonlinearity is clearly seen -- for given physical parameters \cite{Heimburg2005} the smaller solitons move faster than the larger ones \cite{Tamm2015a,Peets2015} . Such an effect differs from the behaviour of soliton trains in the classical Boussinesq equation \cite{Engelbrecht2011}.

\subsection{Pressure waves inside cylindrical biomembranes}
The axoplasm within a nerve fibre is actually a gel consisting 87\% of water held together by cytoskeleton \cite{Gilbert1975}. In modelling such a complicated structure is taken as a pseudoplastic fluid \cite{Biondi1972} or as a viscous compressible fluid \cite{Rvachev2010,Hady2015}. The possible intra-axonal transport of substances in the fluid is not taken into account \cite{Weiss1972}. It means that a pressure wave can be described by Navier-Stokes equations and in the first approximation by the 1D model while the biomembrane can be treated as an elastic tube. In this sense there is a similarity between blood flow in aorta \cite[\textit{etc.}]{Pedley1980,Moodie1983a,Bessems2007} and pressure waves in intersticial fluid.

The governing equation for 1D pressure waves is a momentum balance \cite{Tritton1988}. In terms of longitudinal velocity $v$ it reads: 
\begin{equation}
\label{eq13}
\rho\left(\frac{\pp v}{\pp t}+v\frac{\pp v}{\pp x}\right)+\frac{\pp p}{\pp x}-\mu\frac{\pp^2v}{\pp x^2}=F,
\end{equation}
where $\rho$ is the density, $p$ is the pressure, $\mu$ is the viscosity and F - the body force, and $v$ is the velocity.

In terms of pressure in the 2D setting for waves in fluid surrounded by a cylindrical tube (shell) the governing equations are \cite{Lin1956}:
\begin{subequations}
\begin{align}
&\frac{\pp^2p}{\pp t^2}=c_f^2\left(\frac{\pp^2p}{\pp x^2}+\frac{\pp^2p}{\pp r^2}+\frac{1}{r}\frac{\pp p}{\pp r}\right), \\
&\rho\frac{\pp^2 U}{\pp t^2}+\frac{\pp p}{\pp x}=0,\\
&\rho\frac{\pp^2 V}{\pp t^2}+\frac{\pp p}{\pp r}=0,
\end{align}
\end{subequations}
where $x$ and $r$ are cylindrical coordinates, $c_f$ is the velocity and $U$, $V$ are longitudinal and transverse displacements, respectively. 

\subsection{Coupling}
The ideas of coupling and interrelation of processes related to nerve pulse propagation have been mentioned already long time ago \cite[\textit{etc.}]{Wilke1912,Hodgkin1945}. The recent studies have explained several mechanisms responsible for this complicated physiological phenomenon. The voltage-induced changes in membrane tension and mechanically sensitive ion channels are described in \cite{Mueller2014}. It has been shown how the electric field can generate tension and bending in a membrane which is itself described by a fluid model \cite{Lacoste2007}. Concerning directly the nerve fibres, there is an understanding that an action potential in the nerve fibre generates an axoplasmic pressure wave in the intersticial fluid which propagates synchronously with the action potential \cite{Rvachev2010,Hady2015}. Both can in principle activate the ion channels in the surrounding biomembrane as a result of mechano-electrical activation which plays a crucial role in transmitting the signals. However, this process is not completely understood as well as the coupling of action potential and the pressure wave although many elements of such processes are analysed (Heimburg and Jackson \cite{Heimburg2005}, Rvachev \cite{Rvachev2010}, El Hady and Machta \cite{Hady2015}, etc) together with experimental evidence (Tasaki \cite{Tasaki1988}, Iwasa et al \cite{Iwasa1980}, Shrivastava and Schneider \cite{Shrivastava2014}, etc). It is mentioned in \cite{Vargas2011} that the channel effects usually attributed to action potential are indistinguishable from effects caused by localised pulses in the biomembrane which undergoes phase changes \cite{Mueller2014}. One idea is to describe the coupling by designating the potential energy for a whole system in the biomembrane and the kinetic energy in the movement of the axoplasmic fluid \cite{Hady2015}. Following this model, there are surface modes in the biomembrane driven by changes in charges (electrical pulse) which generate compressive forces on the membrane. The analysis is carried out by using Fourier transform. Rvachev \cite{Rvachev2010} focuses on estimating the velocities of an action potential and the pressure wave driven by it without making an assumption for separate components responsible for kinetic and potential energy like in \cite{Hady2015}.

This is a fundamental question how to model coupling of all processes in a nerve fibre. Clearly a simple combination of action potential models (like HH model) and the models of mechanical waves in biomembranes is not possible. The signal transmittance in neural systems is a complex process with many constituents and many coupling links reflecting the complexity of physiology \cite{Debanne2011,Contreras2013}.
Based on the analysis described above, we shall propose a coupled model based on following assumptions:
 (i) an action potential will generate a pressure wave; (ii) both electric field and mechanical impact can generate opening the ion channels in the membrane; (iii) in the cylindrical biomembrane which surrounds the intersticial fluid, mechanical waves are initiated synchronously with wave processes in the fluid; (iv) biomembrane can exist either in a fluid or in gel state. In the next Section these phenomena will be analysed within one fully coupled model. 

\section{Fully coupled model}

The key elements in modelling seem to be a biomembrane and ion channels which actually govern the amplification of a nerve pulse (an action potential). The simulation of biomembranes up to now is not possible with fully molecular setting \cite{Brown2008}. That is why it is proposed to use continuum mechanics approach and introduce forces for modelling possible coupling \cite{Brown2008}.

The simplest idea is to model nerve pulse by one of the existing action potential models (either by the HH or by the simpler FHN model), the pressure wave in the intersticial fluid by using the Navier-Stokes description and the waves in the cylindrical biomembrane by the improved Heimburg-Jackson model. We base our modelling on the following ideas:

(i) electrical signals are the carriers of information \cite{Debanne2011} and trigger all the other processes;

(ii) the axoplasm in a fibre can be modelled as fluid where a pressure wave is generated due to electrical signal; here, for example, the actin filaments in the axoplasm may influence the opening of channels in the surrounding biomembrane but do not influence the  generation of pressure wave in the fluid \cite{Gilbert1975,Biondi1972,Hady2015} ;

(iii) the biomembrane is able to deform  (stretch, bending) under mechanical impact \cite{Griesbauer2012,Shrivastava2014};

(iv) the channels in biomembranes can be opened and closed under the influence of electrical signals as well as of mechanical input; it means that tension of a membrane leads to the increase of transmembranal ion flow and the intracellular actin filaments may influence the motions at the membrane \cite{Mueller2014,Barz2013}.

(v) there is strong experimental evidence on electrical or chemical transmittance of signals from one neuron to another \cite{Bennett2000,Hormuzdi2004}.

The crucial problem to be solved is how to model the coupling between the action potential, pressure wave, deformation of the biomembrane and opening/closing the ion channels. At this point we leave the possibility of transfer the water molecules through the membrane aside.

\begin{figure}[h]
\includegraphics[height=7cm]{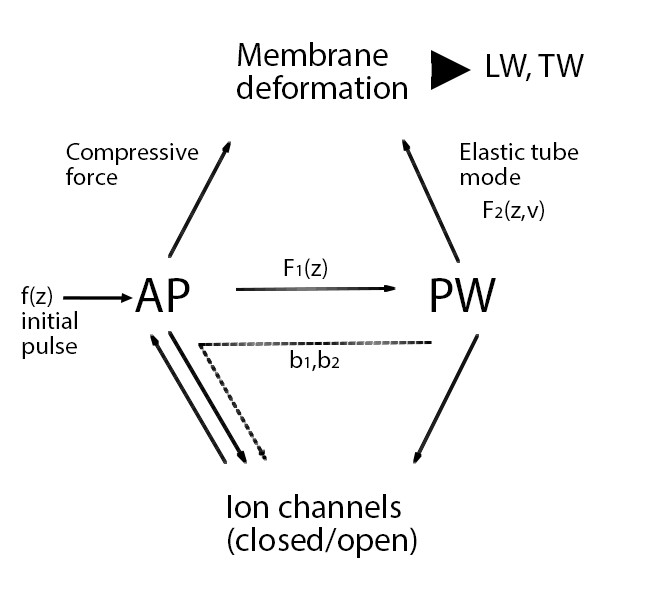}\quad\quad\quad\quad\quad\quad \includegraphics[height=7cm]{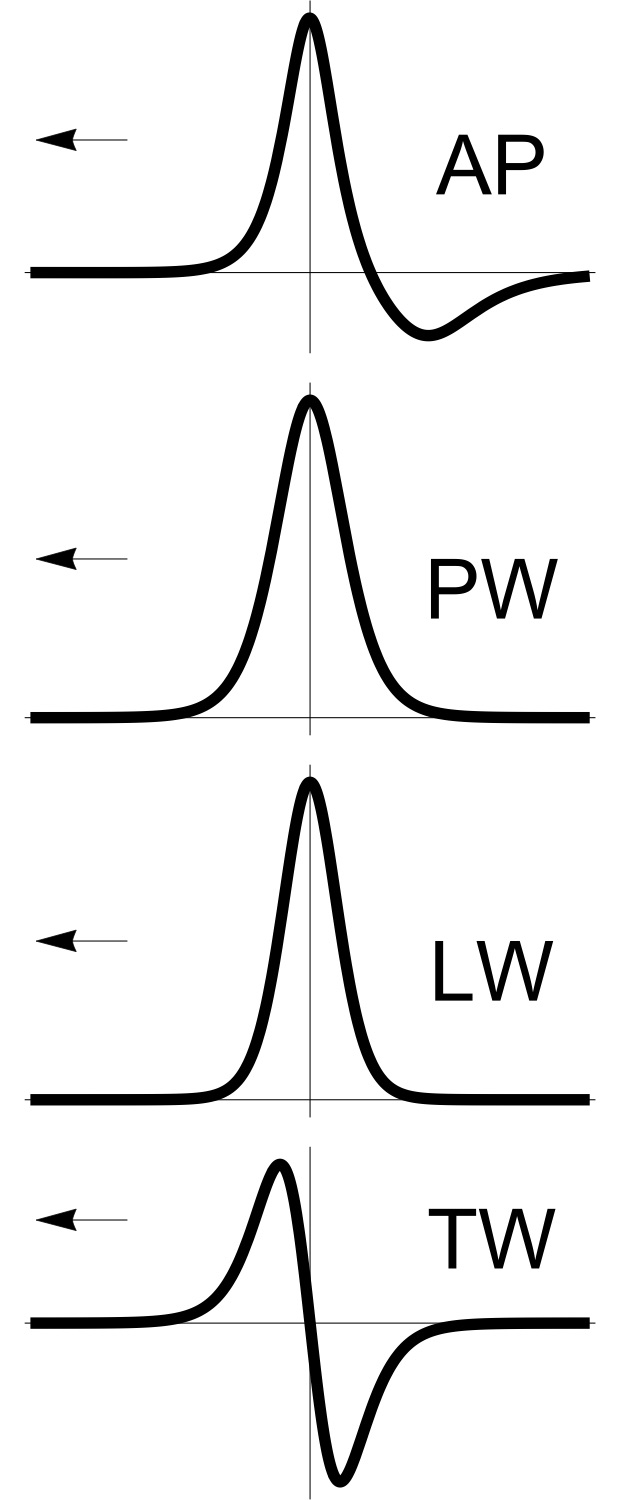}
\caption{The flowchart of coupled models (left) and schemes of waves (right): AP - action potential PW - pressure wave in axoplasm, LW - longitudinal wave in the biomembrane (BM), TW - transverse wave in the BM. Scales are arbitrary.}
\label{Fig1}
\end{figure}

The flowchart of the coupled models is shown and the possible wave profiles are schematically depicted in Fig. \ref{Fig1}. The proposed set of coupled equations is the following:
\\Initial condition:
\begin{equation}
\label{eq15}
z|_{t=0}=f(x),
\end{equation}
where $z$ is an electrical pulse. The action potential is governed by a FHN model:
\begin{equation}
\label{eq16}
\frac{\pp^3 z}{\pp t \pp x^2}=\frac{\pp^2 z}{\pp t^2}+\mu(1-(a_1+b_1)z+(a_2+b_2)z^2)\frac{\pp z}{\pp t}+z,
\end{equation}
where $a_1$, $a_2$ reflect electrical activation and $b_1(v)$, $b_2(v)$ - mechanical activation. The pressure wave in axoplasm is governed by a Navier-Stokes model:
\begin{equation}
\label{eq17}
\rho\left(\frac{\pp v}{\pp t}+v\frac{\pp v}{\pp x}\right)=-\frac{\pp p}{\pp x}+\mu_\nu \frac{\pp^2 v}{\pp x^2}+F_1(z),
\end{equation}
where $v$ is the velocity, $\rho$ the density, $\mu_\nu$ - the viscosity and $F_1(z)$ is a force from the action potential. In the biomembrane, the longitudinal wave is governed by:
\begin{equation}
\label{eq18}
\frac{\pp^2 u}{\pp t^2}=\frac{\pp }{\pp x}\left[\left(c_0^2+pu+qu^2\right)\frac{\pp u}{\pp x}\right]-h_1\frac{\pp^4u}{\pp x^4}+h_2\frac{\pp^4u}{\pp x^2\pp t^2}+F_2(z,v),
\end{equation}
where the notations follow Eqs. \eqref{eq8}, \eqref{eq9} and $F_2(z,v)$ is a force from the process in the axoplasm. In the biomembrane the transverse wave is governed by 
\begin{equation}
\label{eq19}
w=-kr\frac{\pp u}{\pp x},
\end{equation}
where $r=a$ is the axon radius and $k$ is the coefficient which in the theory of rods is the Poisson ratio but in the present case dues to the complicated structure of a biomembrane, needs to be determined from experiments.

\section{Discussion}

The coupled model \eqref{eq16} -- \eqref{eq19} is a set of partial differential equations with coupling forces $F_1(z)$, $F_2(z,v)$. The process is generated by an electrical impulse \eqref{eq15}. Note that according to the soliton model \cite{Heimburg2005} the starting point is a mechanical stimulus which can change the electrical potential and the charge of the membrane \cite{Appali2012a}. Although all the basic single models are well analysed and understood, the coupling forces lead to many open questions. Starting with an input for the action potential, due to the coupling (reciprocity) it should also be possible to model the influence of a mechanical wave in a biomembrane to the action potential. Especially important is how electrical and/or mechanical waves either separately or jointly influence the opening of ion channels \cite{Mueller2014}. Actually one could model this influence either by adjusting nonlinear parameters $a_i$ , $b_i$  in Eq. \eqref{eq16} or by an additional external force $F_3(u,v)$ which acts on the initial system. In the case of the FHN model \eqref{eq16} this assumption means that $a_i + b_i = const$ and an additional term like 
$k_0 \frac{\partial}{\partial t} F_3(u,v)$, $k_0= const$ appears on the r.h.s. of Eq. \eqref{eq16}. Note that Eqs. \eqref{eq16}, \eqref{eq18} are conservative and Eq. \eqref{eq17} includes possible dissipation due to viscosity. 
The possible heat changes are not included into the model but it is possible to account them by using the concept of internal variables \cite{Berezovski2011}. Another important question is to specify the physiological constants which certainly can vary over a large scale \cite{Debanne2011}.
It must be noted that the proposed mathematical model is a robust one which links the basic wave phenomena. The action potential within this fully coupled model is described by the FHN model, which takes only one ion current into account. Certainly, the more detailed HH model can be used. It is not clear yet how the overshoot of an action potential influences the mechanical wave in a biomembrane. Even more, there are several specific phenomena like the variation of ion channel densities, the extracellular accumulation of of ions, the movement of mitochondria, etc which all affect the action potential propagation \cite{Bakkum2013}. 

The crucial problem in modelling is to determine the coupling forces between the physical processes (waves). It must be noted that the present model describes the coupling of variables at a given time moment but the coupling could also be time-dependent taking into account the history of the process. In this case the forces should be described by convulution-type integrals. Here the `history' in some sense is related to the overshoot of an action potential which keeps the sequence of pulses separated by the diffraction length. It is proposed that as far as the mean area of density of the biomembrane stays constant then the compressed regions (solitary wave) should be followed by regions of negative density change \cite{Vargas2011} and this explains the minimum distance between solitary waves.

Another important problem is related to thermodynamics of the full process \cite{Andersen2009}. In order to deform the membrane, a certain amount of work must be done which can be related to the free energy density like Eq. \eqref{eq8}. The general theory based on continuum mechanics derived for biomembranes is given in \cite{Lomholt2006,Deseri2013}, in terms of Canham-Helfrich free energy in \cite{Chou2002} and more specifically related to nerve pulse propagation in \cite{Andersen2009,Heimburg2005,Vargas2011}.It might be useful to explain the work done for deformation directly to energy function like it is done for cardiac mechanoenergetics \cite{Kalda2013} because this might give new insight to the phase transition in the biomembrane. 

Clearly, future studies are needed especially for clarifying the coupling processes. The recording techniques \cite{Hormuzdi2004,Debanne2011,Sasaki2013} developed for studying axon physiology are promising for detecting the characteristics of processes with great accuracy. This is needed not only for determining the parameters but also for understanding the physiological processes. For example, in most studies it is accepted that that an action potential is generated by a synaptic input. In recent studies it is reported that an action potential can be initiated also from the distal axonal region \cite{Sasaki2013}. Due to possible coupling at various levels, the model described in Section 3, might allow such a phenomenon.
The authors hope that the modelling principles described above could help understanding this complicated physiological process.

\section*{Acknowledgements}
This research was supported by the European Union through the European Regional Development Fund (Estonian Programme TK 124) and by the Estonian Research Council (projects IUT 33-7, IUT 33-24)


\begin{thebibliography}{10}

\bibitem{Hodgkin1952}
A.~L. Hodgkin and A.~F. Huxley.
\newblock {A quantitative description of membrane current and its application
  to conduction and excitation in nerve}.
\newblock {\em J. Physiol.}, 117(4):500--544, 1952.

\bibitem{FitzHugh1961}
R.~Fitzhugh.
\newblock {Impulses and physiological states in theoretical models of nerve
  membrane}.
\newblock {\em Biophys. J.}, 1(6):445--466, 1961.

\bibitem{Engelbrecht1981}
J.~Engelbrecht.
\newblock {On theory of pulse transmission in a nerve fibre}.
\newblock {\em Proc. R. Soc. London}, 375(1761):195--209, 1981.

\bibitem{Bresslof2014}
P.~C. Bresslof.
\newblock {\em {Waves in Neural Media}}.
\newblock Springer, 2014.

\bibitem{Tasaki1988}
I.~Tasaki.
\newblock {A macromolecular approach to excitation phenomena: mechanical and
  thermal changes in nerve during excitation}.
\newblock {\em Physiol. Chem. Phys. Med. NMR}, 20:251--268, 1988.

\bibitem{Iwasa1980}
K.~Iwasa, I.~Tasaki, and R.~Gibbons.
\newblock {Swelling of nerve fibers associated with action potentials}.
\newblock {\em Science}, 210(4467):338--339, 1980.

\bibitem{Mueller2014}
J.~K. Mueller and W.~J. Tyler.
\newblock {A quantitative overview of biophysical forces impinging on neural
  function.}
\newblock {\em Phys. Biol.}, 11(5):051001, 2014.

\bibitem{Maugin1994}
G.~A. Maugin and J.~Engelbrecht.
\newblock {A thermodynamical viewpoint on nerve pulse dynamics}.
\newblock {\em J. Non-Equilibrium Thermodyn.}, 19(1), 1994.

\bibitem{Tieleman2003}
D.~P. Tieleman, H.~Leontiadou, A.~E. Mark, and S.-J. Marrink.
\newblock {Simulation of pore formation in lipid bilayers by mechanical stress
  and electric fields.}
\newblock {\em J. Am. Chem. Soc.}, 125(21):6382--3, 2003.

\bibitem{Barz2013}
H.~Barz, A.~Schreiber, and U.~Barz.
\newblock {Impulses and pressure waves cause excitement and conduction in the
  nervous system.}
\newblock {\em Med. Hypotheses}, 81(5):768--72, 2013.

\bibitem{Heimburg1998}
T.~Heimburg.
\newblock {Mechanical aspects of membrane thermodynamics. Estimation of the
  mechanical properties of lipid membranes close to the chain melting
  transition from calorimetry.}
\newblock {\em Biochim. Biophys. Acta}, 1415(1):147--62, 1998.

\bibitem{Rvachev2010}
M.~M. Rvachev.
\newblock {On axoplasmic pressure waves and their possible role in nerve
  impulse propagation}.
\newblock {\em Biophys. Rev. Lett.}, 5(2):73--88, 2010.

\bibitem{Andersen2009}
S.~S.~L. Andersen, A.~D. Jackson, and T.~Heimburg.
\newblock {Towards a thermodynamic theory of nerve pulse propagation.}
\newblock {\em Prog. Neurobiol.}, 88(2):104--13, 2009.

\bibitem{Debanne2011}
D.~Debanne, E.~Campanac, A.~Bialowas, E.~Carlier, and G.~Alcaraz.
\newblock {Axon physiology}.
\newblock {\em Physiol. Rev.}, 91(2):555--602, 2011.

\bibitem{Hormuzdi2004}
S.~G. Hormuzdi, M.~A. Filippov, G.~Mitropoulou, H.~Monyer, and R.~Bruzzone.
\newblock {Electrical synapses: A dynamic signaling system that shapes the
  activity of neuronal networks}.
\newblock {\em Biochim. Biophys. Acta - Biomembr.}, 1662(1-2):113--137, 2004.

\bibitem{Kim2007}
G.~H. Kim, P.~Kosterin, a.~L. Obaid, and B.~M. Salzberg.
\newblock {A mechanical spike accompanies the action potential in Mammalian
  nerve terminals.}
\newblock {\em Biophys. J.}, 92(9):3122--9, 2007.

\bibitem{Bennett2000}
M.~V.~L. Bennett.
\newblock {Electrical synapses, a personal perspective (or history)}.
\newblock {\em Brain Res. Rev.}, 32(1):16--28, 2000.

\bibitem{Zhao2013}
Q.~Zhao.
\newblock {A molecular and biophysical model of the biosignal}.
\newblock {\em Quantum Matter}, 2(1):9--16, 2013.

\bibitem{Scott1999}
A.~Scott.
\newblock {\em {Nonlinear Science. Emergence {\&} Dynamics of Coherent
  Structures.}}
\newblock Oxford University Press, 1999.

\bibitem{Nagumo1962}
J.~Nagumo, S.~Arimoto, and S.~Yoshizawa.
\newblock {An active pulse transmission line simulating nerve axon}.
\newblock {\em Proc. IRE}, 50(10):2061--2070, 1962.

\bibitem{Courtemanche1998}
M.~Courtemanche, R.~J. Ramirez, and S.~Nattel.
\newblock {Ionic mechanisms underlying human atrial action potential properties
  : insights from a mathematical model}.
\newblock {\em Am. J. Physiol.}, 275(1):H301--H321, 1998.

\bibitem{Bean2007}
B.~P. Bean.
\newblock {The action potential in mammalian central neurons.}
\newblock {\em Nat. Rev. Neurosci.}, 8(6):451--65, 2007.

\bibitem{Hodgkin1945}
A.~L. Hodgkin and A.~F. Huxley.
\newblock {Resting and action potentials in single nerve fibres}.
\newblock {\em J. Physiol.}, 104:176--195, 1945.

\bibitem{Appali2012a}
R.~Appali, U.~{Van Rienen}, and T.~Heimburg.
\newblock {A comparison of the Hodgkin-Huxley model and the soliton theory for
  the action potential in nerves}.
\newblock In A.~Iglic, ed., {\em Adv. Planar Lipid Bilayers Liposomes, Vol.
  16}, 275--299. Academic Press, 2012.

\bibitem{Schrader2002}
W.~Schrader, H.~Ebel, P.~Grabitz, E.~Hanke, T.~Heimburg, M.~Hoeckel, M.~Kahle,
  F.~Wente, and U.~Kaatze.
\newblock {Compressibility of lipid mixtures studied by calorimetry and
  ultrasonic velocity measurements}.
\newblock {\em J. Phys. Chem. B}, 106(25):6581--6586, 2002.

\bibitem{Heimburg2005}
T.~Heimburg and A.~D. Jackson.
\newblock {On soliton propagation in biomembranes and nerves.}
\newblock {\em Proc. Natl. Acad. Sci. USA}, 102(28):9790--5, 2005.

\bibitem{Shillcock2002}
J.~C. Shillcock and R.~Lipowsky.
\newblock {Equilibrium structure and lateral stress distribution of amphiphilic
  bilayers from dissipative particle dynamics simulations}.
\newblock {\em J. Chem. Phys.}, 117(10):5048--5061, 2002.

\bibitem{Bockmann2008}
R.~A. B{\"{o}}ckmann, B.~L. de~Groot, S.~Kakorin, E.~Neumann, and
  H.~Grubm{\"{u}}ller.
\newblock {Kinetics, statistics, and energetics of lipid membrane
  electroporation studied by molecular dynamics simulations.}
\newblock {\em Biophys. J.}, 95(4):1837--50, 2008.

\bibitem{Griesbauer2012}
J.~Griesbauer, S.~B{\"{o}}ssinger, A.~Wixforth, and M.~F. Schneider.
\newblock {Propagation of 2D pressure pulses in lipid monolayers and its
  possible implications for biology}.
\newblock {\em Phys. Rev. Lett.}, 108(19):198103, 2012.

\bibitem{Shrivastava2014}
S.~Shrivastava and M.~F. Schneider.
\newblock {Evidence for two-dimensional solitary sound waves in a lipid
  controlled interface and its implications for biological signalling.}
\newblock {\em J. R. Soc. Interface}, 11(97):20140098, 2014.

\bibitem{Petrov2001}
A.~G. Petrov.
\newblock {Flexoelectricity of model and living membranes}.
\newblock {\em Biochim. Biophys. Acta}, 1561:1--25, 2001.

\bibitem{Engelbrecht2015}
J.~Engelbrecht, K.~Tamm, and T.~Peets.
\newblock {On mathematical modelling of solitary pulses in cylindrical
  biomembranes}.
\newblock {\em Biomech. Model. Mechanobiol.}, 14:159--167, 2015.

\bibitem{Hady2015}
A.~{El Hady} and B.~B. Machta.
\newblock {Mechanical surface waves accompany action potential propagation}.
\newblock {\em Nat. Commun.}, 6:6697, 2015.

\bibitem{Appali2012}
R.~Appali, B.~Lautrup, T.~Heimburg, and U.~{Van Rienen}.
\newblock {Soliton collision in biomembranes and nerves-a stability study}.
\newblock {\em J. Math. Ind.}, 16:205--212, 2012.

\bibitem{Lautrup2011}
B.~Lautrup, R.~Appali, A.~D. Jackson, and T.~Heimburg.
\newblock {The stability of solitons in biomembranes and nerves.}
\newblock {\em Eur. Phys. J. E. Soft Matter}, 34(6):1--9, 2011.

\bibitem{Peets2015a}
T.~Peets, K.~Tamm, and J.~Engelbrecht.
\newblock {Numerical investigation of mechanical waves in biomembranes}.
\newblock In S.~Elgeti and J.-W. Simon, editors, {\em Conf. Proc. YIC GACM 2015
  3rd ECCOMAS Young Investig. Conf. 6th GACM Colloquium, July 20-23, 2015,
  Aachen, Ger.}, pages 1--4, 2015.

\bibitem{Gonzalez-Perez2014}
A.~Gonzalez-Perez, R.~Budvytyte, L.~D. Mosgaard, S.~Nissen, and T.~Heimburg.
\newblock {Penetration of action potentials during collision in the median and
  lateral giant axons of invertebrates}.
\newblock {\em Phys. Rev. X}, 4(3):031047, 2014.

\bibitem{Christov2007}
C.~I. Christov, G.~A. Maugin, and A.~V. Porubov.
\newblock {On Boussinesq's paradigm in nonlinear wave propagation}.
\newblock {\em Comptes Rendus M{\'{e}}canique}, 335(9-10):521--535, 2007.

\bibitem{Tamm2015a}
K.~Tamm and T.~Peets.
\newblock {On solitary waves in case of amplitude-dependent nonlinearity}.
\newblock {\em Chaos, Solitons {\&} Fractals}, 73:108--114, 2015.

\bibitem{Peets2015}
T.~Peets and K.~Tamm.
\newblock {On mechanical aspects of nerve pulse propagation and the Boussinesq
  paradigm}.
\newblock {\em Proc. Est. Acad. Sci.}, 64(3S):331--337, 2015.

\bibitem{Engelbrecht2011}
J.~Engelbrecht, A.~Salupere, and K.~Tamm.
\newblock {Waves in microstructured solids and the Boussinesq paradigm}.
\newblock {\em Wave Motion}, 48(8):717--726, 2011.

\bibitem{Gilbert1975}
D.~S. Gilbert.
\newblock {Axoplasm architecture and physical properties as seen in the
  Myxicola giant axon}.
\newblock {\em J. Physiol.}, 253:257--301, 1975.

\bibitem{Biondi1972}
R.~Biondi, M.~Levy, and P.~Weiss.
\newblock {An engineering study of the peristaltic drive of axonal flow}.
\newblock {\em Proc Natl Acad Sci U S A}, 69(7):1732--1736, 1972.

\bibitem{Weiss1972}
P.~A. Weiss.
\newblock {Neuronal dynamics and axonal flow: axonal peristalsis}.
\newblock {\em Proc. Natl. Acad. Sci.}, 69(5):1309--1312, 1972.

\bibitem{Pedley1980}
T.~Pedley.
\newblock {\em {The Fluid Mechanics of Large Blood Vessels}}.
\newblock Cambridge University Press, 1980.

\bibitem{Moodie1983a}
T.~Moodie, D.~Barclay, and R.~Tait.
\newblock {A boundary value problem for fluid-filled viscoelastic tubes}.
\newblock {\em Math. Model.}, 4:195--207, 1983.

\bibitem{Bessems2007}
D.~Bessems, M.~Rutten, and F.~{Van De Vosse}.
\newblock {A wave propagation model of blood flow in large vessels using an
  approximate velocity profile function}.
\newblock {\em J. Fluid Mech.}, 580:145--168, 2007.

\bibitem{Tritton1988}
J.~Tritton.
\newblock {\em {Physical Fluid Dynamics}}.
\newblock Oxford Sci. Publ., 1988.

\bibitem{Lin1956}
T.~Lin and G.~Morgan.
\newblock {Wave propagation through fluid contained in a cylindrical elastic
  shell}.
\newblock {\em J. Acoust. Soc. Am.}, 28(6):1165--1176, 1956.

\bibitem{Wilke1912}
E.~Wilke.
\newblock {On the problem of nerve excitation in the light of the theory of
  waves}.
\newblock {\em Pfl{\"{u}}gers Arch.}, 144:35--38, 1912.

\bibitem{Lacoste2007}
D.~Lacoste, M.~C. Lagomarsino, and J.~F. Joanny.
\newblock {Fluctuations of a driven membrane in an electrolyte}.
\newblock {\em Europhys. Lett.}, 77(1):18006, 2007.

\bibitem{Vargas2011}
E.~V. Vargas, A.~Ludu, R.~Hustert, P.~Gumrich, A.~D. Jackson, and T.~Heimburg.
\newblock {Periodic solutions and refractory periods in the soliton theory for
  nerves and the locust femoral nerve.}
\newblock {\em Biophys. Chem.}, 153(2-3):159--67, 2011.

\bibitem{Contreras2013}
F.~Contreras, H.~Cervantes, M.~Aguero, and M.~d.~L. Najera.
\newblock {Classic and non-classic soliton like structures for traveling nerve
  pulses}.
\newblock {\em Int. J. Mod. Nonlinear Theory Appl.}, 2:7--13, 2013.

\bibitem{Brown2008}
F.~L.~H. Brown.
\newblock {Elastic modeling of biomembranes and lipid bilayers.}
\newblock {\em Annu. Rev. Phys. Chem.}, 59:685--712, 2008.

\bibitem{Berezovski2011}
A.~Berezovski, J.~Engelbrecht, and G.~A. Maugin.
\newblock {Thermoelasticity with dual internal variables}.
\newblock {\em J. Therm. Stress.}, 34(5-6):413--430, 2011.

\bibitem{Bakkum2013}
D.~J. Bakkum, U.~Frey, M.~Radivojevic, T.~L. Russell, J.~M{\"{u}}ller,
  M.~Fiscella, H.~Takahashi, and A.~Hierlemann.
\newblock {Tracking axonal action potential propagation on a high-density
  microelectrode array across hundreds of sites}.
\newblock {\em Nat. Commun.}, 4:2181, 2013.

\bibitem{Sasaki2013}
T.~Sasaki.
\newblock {The axon as a unique computational unit in neurons}.
\newblock {\em Neurosci. Res.}, 75(2):83--88, 2013.

\bibitem{Lomholt2006}
M.A.~Lomholt and L.~Miao.
\newblock {Descriptions of membrane mechanics from microscopic and effective two-dimensional perspectives}.
\newblock {\em J. Phys. A: Math. Gen.}, 39:10323--10354, 2006.

\bibitem{Deseri2013}
L.~Deseri and G.~Zurlo.
\newblock {The stretching elasticity of biomembranes determines their line tension and bending rigidity}.
\newblock {\em Biomech. Model. Mechanobiol.}, 12:1233--1242, 2013.

\bibitem{Chou2002}
T.~Chou.
\newblock {Physics of cellular materials: biomembranes [lecture notes]}.
\newblock {Retrieved from http://faculty.biomath.ucla.edu/tchou/pdffiles/lecture3.pdf}, 2002.

\bibitem{Kalda2013}
M.~Kalda, P.~Peterson, J.~Engelbrecht and M.~Vendelin.
\newblock {A cross-bridge model describing the mechanoenergetics of actomyosin interaction}.
\newblock In G.A.~Holzapfel and E.~Kuhl, editors, {\em Computer Models in Biomechanics, Springer Netherlands}, pages 91--102, 2013.


\end{thebibliography}
\end{document}